\documentclass[12pt,english,american]{article}
\usepackage[T1]{fontenc}
\usepackage[latin9]{inputenc}
\usepackage{geometry}
\usepackage{tikz}
\usepackage{color}
\usepackage{babel}
\usepackage{amsthm}
\usepackage{amstext}
\usepackage{amssymb}
\usepackage{enumitem}
\usepackage{amsmath}
\usepackage{simplewick}
\usepackage{wrapfig}
\usepackage[unicode=true,pdfusetitle,
 bookmarks=true,bookmarksnumbered=false,bookmarksopen=false,
 breaklinks=true,pdfborder={0 0 0},backref=false,colorlinks=true]
 {hyperref}
\hypersetup{
 linkcolor=blue,citecolor=blue, urlcolor=blue}
\usepackage{caption}

\makeatletter

\newcommand{\mrm}{\mathrm}
\newcommand{\vjl}{\nabla E_{p_{\ell} } }

\usepackage{enumitem}		
\theoremstyle{plain}

\theoremstyle{definition}

\theoremstyle{plain}

\theoremstyle{remark}

\theoremstyle{plain}

\theoremstyle{plain}

\newcommand{\unp}{\un p}

\usepackage{txfonts,refstyle,xcolor}

\newref{con}{name = Conjecture\ }
\newref{prop}{name = Proposition\ }
\newref{def}{name = Definition\ }
\newref{sec}{name = Section\ }
\newref{sub}{name = Section\ }
\newref{thm}{name = Theorem\ }
\newref{lem}{name = Lemma\ }
\newref{cor}{name = Corollary\ }
\newref{fig}{name = Figure\ }

\usepackage{bbm}

\usepackage[all]{xy}

\newcommand{\xyR}[1]{%
     \makeatletter
     \xydef@\xymatrixrowsep@{#1}
     \makeatother
}

\newcommand{\xyC}[1]{%
     \makeatletter
     \xydef@\xymatrixcolsep@{#1}
     \makeatother
}

\newcommand{\ncol}[1]{\color{normalcolor}}

\makeatother

\newcommand{\Ome}{|0\ran}

\addto\captionsamerican{\renewcommand{\corollaryname}{Corollary}}
\addto\captionsamerican{\renewcommand{\definitionname}{Definition}}
\addto\captionsamerican{\renewcommand{\lemmaname}{Lemma}}
\addto\captionsamerican{\renewcommand{\propositionname}{Proposition}}
\addto\captionsamerican{\renewcommand{\remarkname}{Remark}}
\addto\captionsamerican{\renewcommand{\theoremname}{Theorem}}
\addto\captionsenglish{\renewcommand{\corollaryname}{Corollary}}
\addto\captionsenglish{\renewcommand{\definitionname}{Definition}}
\addto\captionsenglish{\renewcommand{\lemmaname}{Lemma}}
\addto\captionsenglish{\renewcommand{\propositionname}{Proposition}}
\addto\captionsenglish{\renewcommand{\remarkname}{Remark}}
\addto\captionsenglish{\renewcommand{\theoremname}{Theorem}}
\providecommand{\corollaryname}{Corollary}
\providecommand{\definitionname}{Definition}
\providecommand{\lemmaname}{Lemma}
\providecommand{\propositionname}{Proposition}
\providecommand{\remarkname}{Remark}
\providecommand{\theoremname}{Theorem}

\newcommand{\as}{\mathrm{as}}
\newcommand{\vj}{\nabla E_p}
\newcommand{\D}{\mathrm{D}}

\newcommand{\II}{\mathrm{I}}

\newcommand{\ren}{\mathrm{ren}}
\newcommand{\peps}{\ov{\eps}}

\newcommand{\bc}{ \color{black} }
\newcommand{\x}{\!\!\!}

\newcommand{\np}{\ov{n}}

\renewcommand{\S}{\mathrm S}
\newcommand{\W}{\mathcal W}

\newcommand{\hc}{\mathrm{h.c.}}

\newcommand{\unk}{\un k}

\newcommand{\hk}{\hat k }    

\newcommand{\pho}{\mathrm{ph}}

\newcommand{\nn}{\eta}

\newcommand{\bb}{a}

\newcommand{\vv}{v}

\newcommand{\ti}{\tilde}

\newcommand{\un}{\underline}

\newcommand{\Om}{\Omega}
\newcommand{\ga}{\gamma}

\newcommand{\ka}{\kappa}

\newcommand{\be}{\beta}

\newcommand{\ov}{\overline}

\newcommand{\eps}{\varepsilon}

\newcommand{\e}{e}

\newcommand{\nin}{\noindent}
\newcommand{\si}{\sigma}
\newcommand{\ph}{\phantom}
\newcommand{\h}{\fr{1}{2}}
\newcommand{\nat}{\mathbb{N}}
\newcommand{\hil}{\mathcal{H}}
\newcommand{\om}{\omega}

\newcommand{\fr}[2]{\frac{#1}{#2}}
\newcommand{\al}{\alpha}
\newcommand{\real}{\mathbb{R}}

\newcommand{\la}{\lambda}
\newcommand{\non}{\nonumber}
\newcommand{\Ga}{\Gamma}

\newcommand{\ran}{\rangle}
\renewcommand{\eta}{b}

\newtheorem{theoreme}{Theorem } [section]
\newtheorem{proposition}[theoreme]{Proposition}
\newtheorem{lemma}[theoreme]{Lemma}
\newtheorem{definition}[theoreme]{Definition}
\newtheorem{corollary}[theoreme]{Corollary}
\newtheorem{remark}[theoreme]{Remark}
\newtheorem{example}[theoreme]{Example}
\newtheorem{criterion}[theoreme]{Criterion}

\newcommand{\beq}{\begin{equation}}
\newcommand{\eeq}{\end{equation}}
\newcommand{\beqa}{\begin{eqnarray}}
\newcommand{\eeqa}{\end{eqnarray}}
\newcommand{\ben}{\begin{arabicenumerate}}
\newcommand{\een}{\end{arabicenumerate}}
\newcommand{\bex}{\begin{example}}
\newcommand{\eex}{\end{example}}
\newcommand{\ber}{\begin{remark}}
\newcommand{\eer}{\end{remark}}
\newcommand{\bec}{\begin{corollary}}
\newcommand{\eec}{\end{corollary}}
\newcommand{\bep}{\begin{proposition}}
\newcommand{\eep}{\end{proposition}}
\newcommand{\becr}{\begin{criterion}}
\newcommand{\eecr}{\end{criterion}}

\newcommand{\hatk}{\hat{k}}
\newcommand{\Omm}{\om}
\newcommand{\mcF}{\mathcal{F}}
\renewcommand{\e}{\mathrm{e}}
\newcommand{\fiber}{\mathrm{fi}}

\def\bel{\begin{lemma}}
\def\eel{\end{lemma}}
\def\bet{\begin{theoreme}}
\def\eet{\end{theoreme}}
\def\bed{\begin{definition}}
\def\eed{\end{definition}}

\begin{document}
\title{From Faddeev-Kulish to  LSZ. \\ Towards a non-perturbative description of colliding electrons } 

\author{
{\bf Wojciech Dybalski}\\
Zentrum Mathematik, Technische Universit\"at M\"unchen,\\
and\\
Fakult\"at f\"ur Mathematik, Ludwig-Maximilians-Universit\"at M\"unchen \\
E-mail: {\tt dybalski@ma.tum.de}
}

\date{}

\maketitle

\vspace{-0.7cm}
\begin{center}
\small{\emph{Dedicated to the memory of Wolfhart Zimmermann}}
\end{center}
\vspace{0.1cm}
\begin{abstract}  In a low energy approximation of the massless Yukawa theory (Nelson model) we derive a Faddeev-Kulish type 
formula for the scattering matrix of $N$ electrons and reformulate it in  LSZ terms. To this end, we perform 
a decomposition of the  infrared finite Dollard modifier into clouds of real and virtual  photons, 
whose infrared divergencies mutually cancel. 
We point out that in the original work of Faddeev and Kulish the  clouds of real  photons are omitted, 
 and consequently their scattering matrix  is ill-defined on the Fock space of free electrons. To support our observations,
we compare our final LSZ expression for $N=1$ with a rigorous non-perturbative construction 
 due to Pizzo.  While our  discussion contains some heuristic steps, 
they can be formulated as  clear-cut mathematical conjectures.


\end{abstract}

\section{Introduction}\setcounter{equation}{0}
Infrared problems enjoyed recently a revival, triggered by 
works of Strominger et al. on relations between soft photon theorems, asymptotic
symmetries and memory effects  (see \cite{St17} for a review).  One line of  developments consisted
in reformulating this `infrared triangle' in terms of modified asymptotic dynamics in the sense of Faddeev 
and Kulish  \cite{GS16, GP16,GS17,Pa17, MP16, KPRS17}.  Given the ambitions of these recent advances,
reaching quantum gravity and black-hole physics, we have to point out that  the  mathematical and conceptual basis of the 
Faddeev-Kulish approach is not very solid, not even in its original context. First of all, both in the original work \cite{FK70} and in the recent references, the Faddeev-Kulish approach is 
justified at best by working out some test cases in perturbation theory.  The question  if the infrared finite $S$-matrix has any non-perturbative 
meaning is left completely open. Secondly, the relation between the Faddeev-Kulish approach to the more standard LSZ scattering theory has never been
clarified. While a naive application of the LSZ ideas clearly fails in the presence of infrared problems, a careful LSZ description of a bare electron
accompanied by real and virtual photons is in fact possible \cite{Fr73,Pi05, CFP07}.  In the present work we outline a bridge from the Faddeev-Kulish 
formalism to this LSZ description in the massless Nelson model. 

The Nelson model has been used for many decades for non-perturbative discussions of
infrared problems (see e.g. \cite{Fr73, Fr74.1, Pi05, AH12, DP12.0}). Its Hamiltonian, stated  in Section~\ref{The-model} below,  can be obtained as a low energy approximation of the massless Yukawa theory with the interaction Lagrangian $\mathcal{L}_{\mathrm{I}}=\la\ov{\psi} \phi \psi$.   Here $\psi$ is the massive Dirac field, whose excitations will be called electrons/positrons, and $\phi$ is the massless scalar field whose excitations will be called photons (although they are spinless). We fix an ultraviolet cut-off $\ka$ and approximate the dispersion relation of the massive particles by the non-relativistic formula $p\mapsto p^2/(2m)$, where $m=1$ for simplicity.
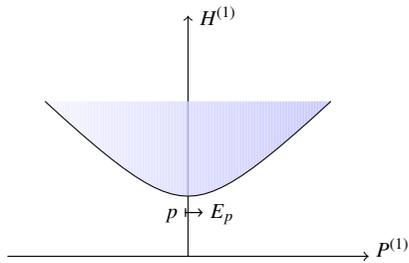
\begin{wrapfigure}{r}{0.5\textwidth}
\begin{center}
\begin{tikzpicture}[scale=0.8]
                        \begin{scope}[->]
            \draw[black] (-3,0) -- (3,0) node[anchor=north] {};
            \draw[black] (0,0) -- (0,4) node[anchor=east] {};
        \end{scope}
       \draw[fill,black] (3.4,.15) node{\scriptsize{$P^{(1)}$}};
       \draw[fill,black] (.5,4) node{\scriptsize{$H^{(1)}$}};
        \draw[fill,black] (0.2,.7) node{\scriptsize{$p\mapsto E_p$}};
               \shade[left color=blue!5!white,right color=blue!30!white,opacity=0.5] plot[variable=\t,domain=-1.6:1.6] ({sinh(\t)},{cosh(\t)});
       \draw plot[variable=\t,domain=-1.6:1.6] ({sinh(\t)},{cosh(\t)});
       \end{tikzpicture}
    \caption*{\small{Figure 1: The energy-momentum spectrum in the single-electron sector of the massless Nelson model.} }    
       \end{center}
 \end{wrapfigure}      
 As the creation and annihilation processes of electron-positron 
pairs can be neglected in the low-energy regime, we can restrict attention to the zero-positron sector and include only the electron-photon interactions in the
Hamiltonian $H$ of the Nelson model. 
This Hamiltonian commutes with the total number of electrons and we denote by $H^{(N)}$ the $N$-electron Hamiltonians.  
Furthermore, by the translation invariance of the model,  $H^{(N)}$ 
commutes with the respective total momentum operator $P^{(N)}$ and thus
this family of operators can be diagonalized simultaneously.  For $N=1$ the lower boundary of their joint spectrum   
is the physical (renormalized) energy-momentum relation of the electron which we denote $p\mapsto E_p$ (see Figure~1). 
This dispersion relation has been a subject of study for many decades and it is relatively well understood \cite{AH12, Fr74.1,Pi03, DP12}.
Two comments about its properties are in order, since they anticipate our discussion in the later part of this paper:
\begin{enumerate}
\item[(a)] In the presence of interaction the physical dispersion relation  $p\mapsto E_p$ differs from the bare one $p\mapsto \fr{p^2}{2}$ appearing
in the free Hamiltonian (\ref{free-Hamiltonian}). This  is caused by certain photon degrees of freedom  `sitting' on  the bare electron, which are responsible, in particular,
for radiative corrections to its mass. We will refer to these photons as `clouds of virtual photons', to distinguish them from `clouds of real photons' described in (b) below. 
In the following discussion these virtual photons will appear in the step from the bare creation operator  $\eta^*(p)$ to the renormalized creation operator
 $\ti \eta_{\si}^*(p)$  of the electron  (cf. formula~(\ref{tentative-rco}) below).

\item[(b)]  It is also well known that there are no normalizable states in the Hilbert space of the model, that would `live' 
exactly at the lower boundary of the spectrum from Figure 1. 
In other words,  it is not possible to find normalizable states  describing just the physical electron (including its cloud of virtual photons) and no other particles. 
Hence, the  electron is always accompanied by some `cloud of real photons', 
moving to lightlike infinity.   This cloud, denoted $\W_{p,\si}(t)$,
 will also appear naturally in our discussion below, see  (\ref{cloud-of-photons}).

\end{enumerate}
An early discussion of the Faddeev-Kulish formalism in the Nelson model is due to Fr\"ohlich~\cite[Chapter 5]{Fr73}, who was quite pessimistic about
 its rigorous mathematical justification. Our work  still contains some heuristic steps, but they have a form of plausible, clear-cut
conjectures (see Sections~\ref{photon-clouds-section} and  \ref{rigorous}).
As one can expect, we  start in Section~\ref{FK-section} below  from the concept of the Dollard modifier $U^{\D}_{\unp}(t)$,
which comes from quantum mechanical long-range scattering. It does not suffer from any infrared divergencies and thus does not require
infrared regularization. 
Such divergencies  appear only in Section~\ref{regularization} when we start rewriting the Faddeev-Kulish scattering states in LSZ terms. This is completed in Section~\ref{photon-clouds-section}, where we express the 
quantity $U^{\D}_{\unp}(t)$ as a product of infrared divergent objects of two types: the  clouds of real photons $\W_{\unp,\si}(t)$ and the renormalized 
creation operators $\ti\eta_{\si}^*(p)$, 
 both of which are  well-defined only in the presence of an infrared cut-off $\si>0$. From this perspective it is completely clear, that the two types of infrared divergencies, discussed in (a) and (b) above, must mutually cancel as $\si\to 0$.  In Section~\ref{rigorous} we  indicate that the resulting LSZ formula in the case $N=1$ 
reproduces, up to  minor technical differences,  a rigorous formula for one-electron scattering states in the Nelson model due to Pizzo~\cite{Pi05}. We conclude our discussion  with several clear-cut mathematical conjectures concerning the convergence of $N$-electron scattering state approximants in the Nelson model.   

Strangely, the original work of Faddeev and Kulish misses the central point above, namely the cancellation of  infrared divergences coming from the clouds of
real and virtual photons. 
In fact,  the omission of the lower boundary of integration  in formula (9) of \cite{FK70} (which corresponds to dropping term~(\ref{lower-boundary}) below)
ensures commutation of the $S$-matrix with the total momentum of  charged particles. Consequently, there is
no room for clouds of real photons and the $S$-matrix is ill-defined on the Fock space of free electron states. Faddeev and
Kulish try to cure this problem by a contrived  construction of the asymptotic Hilbert space, based on singular coherent
states. While this  strategy may work in some test-cases in perturbation theory, to our knowledge it has never matured into a non-perturbative  argument.

Some aspects of this problem have recently been  noticed in \cite{GP16}, but the  modification 
of the Faddeev-Kulish ansatz in this reference is somewhat ad hoc.  Our solution is very natural: we  apply the Dollard  
formalism according to the rules of the art \cite{DG}, without tampering with the lower boundary of integration. The resulting $S$-matrix may not commute with the
total momentum of the electrons, but it acts on the usual Fock space.  
As mentioned above, the resulting scattering state
can be given a solid LSZ interpretation in terms of electrons dressed with clouds of virtual photons and accompanied by clouds of real photons. It should be pointed out, that a similar picture
of the electron is behind the well-tested Yennie-Frautschi-Suura algorithm for inclusive cross-sections \cite{YFS61}.  

\section{The model}\label{The-model}  \setcounter{equation}{0}
 
The Hilbert space of the Nelson model  is given by
 $\hil=\mcF_{\e}\otimes \mcF_{\pho}$, where $\mcF_{\e}$, $\mcF_{\pho}$
are the Fock spaces of the electrons and photons with creation and annihilation operators denoted $\eta^{(*)}$, $a^{(*)}$, respectively.   
The Hamiltonian of this model is given by
\beqa
& &H:=H_0+V, \label{Nelson-model}\\
& &H_{0}:=\int d^3p\, \fr{p^2}{2}\nn^*(p)\nn(p)+\int d^3k\, |k| a^*(k)a(k),\label{free-Hamiltonian}\\
& &V:=\int d^3p d^3k\,\vv(k)\big(\nn^*(p+k)\bb(k)\nn(p)+\hc\big), \quad \vv(k):=\la \fr{\chi_{[0,\kappa]}(|k|) }{ \sqrt{2|k|}  },
\label{full-auxiliary-model}
\eeqa
where $H_0$ involves the free evolution of the electrons and photons, $V$ is the interaction,
$\ka$ is a fixed ultraviolet cut-off and  $\chi_{[0,\ka]}(|k|)=1$ for  $0\leq |k|\leq \ka$ and $ \chi_{[0,\ka]}(|k|)=0$ otherwise.  
As the Fermi statistics and  the spin degrees of freedom of the electron will not play any role in the following discussion, we suppress the latter in the notation. 

Since this Hamiltonian commutes with the total number $N$ of electrons, 
we can consider the Hamiltonians $H^{(N)}$  on the $N$-electron subspace $\hil^{(N)}:=\mcF^{(N)}_{\e}\otimes \mcF_{\pho}$, given by
\beqa
H^{(N)}=\sum_{\ell=1}^{N}\fr{(-i\nabla_{x_{\ell} })^2}{2}+\int d^3k\, |k| a^*(k)a(k)
+\sum_{\ell=1}^{N}\int d^3k\, \vv(k)\, ( e^{ikx_{\ell}}a(k)+e^{-ikx_{\ell}}a^*(k) ), \label{N-electron-Hamiltonian}
\label{explicit-Hamiltonian}
\eeqa
where $x_{\ell}$ is the position operator of the $\ell$-th electron and $\mcF^{(N)}_{\e}$ is the $N$-particle subspace of $\mcF_{\e}$. This quantum-mechanical
representation will facilitate the application of the Dollard prescription in Section~\ref{FK-section}.

\section{The Dollard formalism } \label{FK-section}  \setcounter{equation}{0}

As we are primarily interested in electron collisions, we treat all photons in the model as `soft' and do not introduce
any division of  the range of photon energies $[0,\ka]$ into a soft and hard part.  
Our starting point is the interaction $V$, which is given on $\hil^{(N)}$ by
\beqa
V=\sum_{\ell=1}^{N}\int d^3k\, v(k) \big(e^{-ikx_{\ell}} a^*(k)+e^{ikx_{\ell}} a(k)\big). \label{potential}
\eeqa
According to the Dollard prescription, we construct the
asymptotic interaction as follows: We substitute $x_{\ell}\to\nabla E_{p_{\ell}} t$, where $\nabla E_{p_{\ell}}$ is the velocity of the
$\ell$-th electron moving with momentum $p_{\ell}$  along the ballistic trajectory, as expected for asymptotic times. 
Thus we have
\beqa
V^{\as}_{\unp }(t):=\sum_{\ell=1}^N\int d^3k\, v(k)\, \big(e^{-ik\cdot \nabla E_{p_{\ell}} t} a^*(k)+e^{ik \cdot \nabla E_{p_{\ell}}  t} a(k)\big), \label{asymptotic-potential}
\eeqa
where $\unp:=(p_1,\ldots, p_N)$ are  momenta of the electrons.
As the physical dispersion relation of the electron is not $p\mapsto p^2/2$ appearing in $H_0$ but rather the lower boundary $p\mapsto E_p$ of the energy-momentum spectrum,  we define the renormalized free Hamiltonian:
\beqa
H^{\ren}_{0}:= \int d^3p\,(E_{p}-C_p)\eta^*(p)\eta(p)+\int d^3k \, |k| a^*(k) a(k), \quad C_p:= \int d^3k\,\,\fr{  v(k)^2}{\Om_p(k)}.
\eeqa
Here $\Om_p(k):=|k|- k\cdot \vj$ and the choice of the normalization constant $C_p$ will be justified a posteriori in Section~\ref{photon-clouds-section}.  
(The need to renormalize the free Hamiltonian was noted already in \cite{Fr73}). 
Thus the asymptotic interaction in the interaction picture is
\beqa
V^{\as, \II}_{\unp}(t)=\e^{i H^{\ren}_{0} t}  V^{\as}_{\unp}(t) \e^{-i H^{\ren}_{0} t}\x&=&\x\sum_{\ell=1}^{N}\int d^3k\, v(k)\, \big(e^{i(|k|-k\cdot \vjl) t} a^*(k)+e^{-i(|k|-k\cdot \vjl) t} a(k)\big)\non\\
\x&=&\x\sum_{\ell=1}^{N}\int d^3k\, v(k)\, \big(e^{i\Om_{p_{\ell}}(k) t} a^*(k)+e^{-i \Om_{p_{\ell}}(k) t} a(k)\big).
\eeqa
Now we define the Dollard modifier
\beqa
U^{\D}_{\unp}(t):=\mathrm{T}\exp\big(-i\int_0^td\tau V^{\as,\II}_{\unp}(\tau)\big)=
e^{-i\int_0^td\tau \, V^{\as,\II}_{\unp}(\tau)  -\h \int_0^t d\tau_1\int_0^{\tau_1}d\tau_2 \, [V^{\as, \II}_{\unp}(\tau_1),  V^{\as, \II}_{\unp}(\tau_2) ] }, \label{dollard}
\eeqa
where the second step above is standard \cite{FK70}.
For any  family of functions  $h_{\ell}\in C_0^{\infty}(\real^{3})$, $\ell=1,\ldots, N$, of the electron momenta we define the corresponding scattering state approximant as follows:
\beqa
\Psi_{h,t}\x&=&\x e^{iHt}e^{-iH_0^{\ren} t}\int d^{3}p_1\ldots d^3p_N\, U^{\D}_{\unp}(t) h_1(p_1)\ldots h_N(p_N )\eta^*(p_1)\ldots \eta^*(p_N)\Ome \non\\ 
\x&=&\x e^{iHt} e^{-iH_0^{\ren} t} \int d^{3N}\unp\, U^{\D}_{\unp}(t) h(\unp )\eta^*(\unp)^N\Ome,
\label{scattering-states-approximants-x}
\eeqa
where in the second step we introduced some obvious short-hand notation.
We note that all quantities above  are well defined without infrared regularization. But a need for infrared regularization will arise in the next subsection, where
we start reformulating states (\ref{scattering-states-approximants-x}) in terms of the LSZ asymptotic creation operators of photons and electrons, 
whose approximating sequences are given schematically by
\beqa
t\mapsto e^{iHt} \big(e^{-i|k|t} a^{*}(k) \big) e^{-iHt}, \quad t\mapsto e^{iHt} \big(e^{-iE_{p}t} b^{*}(p)\big) e^{-iHt}.
\eeqa
As we will see in  (\ref{tentative-rco})--(\ref{actual-rco}) below, $b^{*}(p)$ will actually require renormalisation.

To conclude this section, we  define the wave-operators $\Om^{\mathrm{in}/\mathrm{out}}: \mcF_{\mathrm{e}} \to \hil$  for the electron scattering as follows
\beqa
\Om^{\mathrm{in}/\mathrm{out}}\big(\int d^{3N}\unp\, h(\unp )\eta^*(\unp)^N\Ome\big)=\lim_{t\to-/+\infty} e^{iHt} e^{-iH_0^{\ren} t} \int d^{3N}\unp\, U^{\D}_{\unp}(t) h(\unp )\eta^*(\unp)^N\Ome \label{wave-operator}
\eeqa
so that the corresponding scattering matrix $S:=(\Om^{\mathrm{out} })^* \Om^{\mathrm{in} }$ is an operator on $\mcF_{\mathrm{e}}$. The existence of
the limit in (\ref{wave-operator}) is not settled, but seems to be a feasible functional-analytic problem, as we discuss in Section~\ref{rigorous}. 
\section{Infrared regularization}\label{regularization}   \setcounter{equation}{0}

Let us consider the exponential in the Dollard modifier~(\ref{dollard}) and perform the time integral
\beqa
-i\int_0^td\tau\, V^{\as, \II}_{\unp}(\tau)\x &=&\x (-i)\sum_{\ell=1}^{N} \int d^3k\, v(k) \bigg( \fr{(e^{i\Om_{p_{\ell}}(k) t}-1)}{i\Om_{p_{\ell}}(k)} a^*(k)
+\fr{(e^{-i\Om_{p_{\ell}}(k) t}-1 )}{(-i)\Om_{p_{\ell}}(k)}a(k)\bigg) \label{time-integral} \\
\x &=&\x \sum_{\ell=1}^{N}  \int d^3k\, \fr{v(k)}{\Om_{p_{\ell}}(k)} \big( a^*(k)- a(k)\big) \label{lower-boundary}\\
\x & &\x - \sum_{\ell'=1}^{N} \int d^3k\, \fr{v(k)}{\Om_{p_{\ell'}}(k)} \big( e^{i\Om_{p_{\ell'}}(k) t}   a^*(k)- e^{-i\Om_{p_{\ell'}}(k) t} a(k)\big).
\label{upper-boundary}
\eeqa
Since the l.h.s. of (\ref{time-integral}) is manifestly  infrared finite, the same is true for the r.h.s. of this expression. However,  terms (\ref{lower-boundary})
and (\ref{upper-boundary}) considered separately, coming from the lower and upper boundary of the $\tau$-integration, are infrared singular.
Indeed, they involve $a^{(*)}(k)$ integrated with functions which have a non-square-integrable singularity at zero momentum.  This  division of a regular expression into two singular parts, which will be needed to express the approximating vector (\ref{scattering-states-approximants-x}) in the LSZ fashion, is the source of infrared divergencies, which must mutually cancel. As we pointed out above, in the work of Faddeev and Kulish \cite{FK70} the counterpart of (\ref{lower-boundary})
is omitted.

To make sense out of (\ref{lower-boundary}) and (\ref{upper-boundary}), we need some infrared regularization of (\ref{scattering-states-approximants-x}). 
To this end, we introduce an infrared cut-off $\si>0$ and define a regularized version of the form factor from (\ref{full-auxiliary-model})
\beqa
\vv^{\si}(k):=\la \fr{\chi_{[\si,\kappa]}(|k|) }{ \sqrt{2|k|}  }, \label{form-factor-formula}
\eeqa 
where $\chi_{[\si,\kappa]}(|k|)=1$ for $\si \leq |k|\leq \ka$ and $\chi_{[\si,\kappa]}(|k|)=0$ otherwise. The corresponding potential
and Hamiltonians are denoted $V_{\si}$, $H_{\si}$, $H^{(N)}_{\si}$ and $p\mapsto E_{p,\si}$ is the resulting dispersion relation of the electron.
Next, we  define the regularized approximating sequence analogously as in the previous section
\beqa
\Psi^{\si}_{h,t}=e^{iHt}e^{-iH^{\ren}_{0;\si} t}\int d^{3N}\unp\, U^{\D}_{\unp,\si}(t) h(\unp)\eta^*(\unp)^N\Ome, \label{scattering-states-approximants-x-x}
\eeqa
with the help of the regularized quantities:
\beqa
& &H^{\ren}_{0;\si}:= \int d^3p\, (E_{p,\si}-C_{p,\si})\eta^*(p)\eta(p)+\int d^3k \, |k| a^*(k) a(k), \label{ren-Hamiltonian} \\
& &U^{\D}_{\unp,\si}(t):=\mathrm{T}\exp\big(-i\int_0^td\tau\, V^{\as,  \II}_{\unp,\si}(\tau)\big)=e^{-i\int_0^td\tau V^{\as,\II}_{\unp,\si}(\tau)  
-\h \int_0^t d\tau_1\int_0^{\tau_1}d\tau_2[V^{\as,\II}_{\unp,\si}(\tau_1),  V^{\as,\II}_{\unp,\si}(\tau_2) ] }, \label{exponential-map-sigma}\\
& &V^{\as, \II}_{\unp, \si}(t)
=\sum_{\ell=1}^{N}\int d^3k\, v^{\si}(k)\, \big(e^{i\Om_{p_{\ell}, \si}(k) t} a^*(k)+e^{-i \Om_{p_{\ell}, \si}(k) t} a(k)\big), \label{V-as-reg}
\eeqa
where $C_{p,\si}:= \int d^3k\,\,\fr{  v^{\si}(k)^2}{\Om_{p, \si}(k)}$ and $\Om_{p, \si}(k):=|k|- k\cdot \nabla E_{p,\si}$.
In this situation we have, analogously as in (\ref{lower-boundary})--(\ref{upper-boundary}),  
\beqa
-i\int_0^td\tau\, V^{\as,\II}_{\unp,\si}(\tau)\x&=&\x  \sum_{\ell=1}^{N}\int d^3k\, \fr{v^{\si}(k)}{\Om_{p_{\ell},\si}(k)}\, \big( a^*(k)- a(k)\big) \label{A}\\
& &-\sum_{\ell'=1}^{N}\int d^3k\, \fr{v^{\si}(k)}{\Om_{p_{\ell'},\si}(k)}\, \big(e^{i\Om_{p_{\ell'},\si}(k) t} a^*(k)-e^{-i\Om_{p_{\ell'},\si}(k) t} a(k)\big), \label{B}
\eeqa
but the two terms (\ref{A}) and (\ref{B})  above are now well defined and can be analyzed  separately. By a straightforward computation using
the Baker-Campbell-Hausdorff formula, we thus obtain from~(\ref{scattering-states-approximants-x-x})  
\beqa
\Psi^{\si}_{h,t}\x&=&\x e^{iHt}e^{-iH^{\ren}_{0;\si} t}\int d^{3N}\unp\,  e^{i\ga_{\unp,\si}(t)}e^{-\theta_{\unp,\si}(t) } \prod_{\ell=1}^N \bigg( e^{iC_{p_{\ell},\si }t  }e^{\int d^3k\, \fr{v^{\si}(k)}{\Om_{p_{\ell},\si}(k)}\, ( a^*(k)- a(k))} \bigg) \times\non\\
\x & &\x\ph{44444444444444444444444}\times \prod_{\ell'=1}^N \bigg( e^{-D_{p_{\ell'},\si }} e^{-\int d^3k\, \fr{v^{\si}(k)}{\Om_{p_{\ell'},\si}(k)}\, e^{i\Om_{p_{\ell'},\si}(k) t} a^*(k)  } \bigg)  h(\unp)\eta^*(\unp)^N\Ome,\quad\quad \label{approx-vec-after-BCH}
\eeqa
where $C_{p,\si}$ appeared below (\ref{V-as-reg}) and $D_{p,\si}:=\h \int d^3k \fr{v^{\si}(k)^2 }{\Om_{p,\si}(k)^2}$. The real-valued numerical functions $\ga_{\unp,\si}, \theta_{\unp,\si}$ are stated in (\ref{phase-starts})--(\ref{phase-ends}) below and will be discussed  later. 

\section{Clouds of real and virtual photons, phases} \label{photon-clouds-section}   \setcounter{equation}{0}

We now rewrite formula (\ref{approx-vec-after-BCH}) in the LSZ fashion to facilitate its interpretation in terms of real and virtual photon clouds. 
 By shifting the term $e^{-iH^{\ren}_{0;\si} t}$ to the right
and noting the cancellation of the constants  $C_{p_{\ell},\si }$ (cf. (\ref{ren-Hamiltonian})) we get
\beqa
\Psi^{\si}_{h,t}\x&=&\x e^{iHt}\int d^{3N}\unp\,  e^{i\ga_{\unp,\si}(t)}e^{-\theta_{\unp,\si}(t) } \prod_{\ell=1}^N \bigg( e^{\int d^3k\, \fr{v^{\si}(k)}{\Om_{p_{\ell},\si}(k)}\, ( e^{-i|k|t}a^*(k)- e^{i|k|t}a(k))} \bigg) \times\label{LSZ-cloud}\\
& &\ph{44444444444444444}\times \prod_{\ell'=1}^N \bigg( e^{-D_{p_{\ell'},\si }} e^{-\int d^3k\, \fr{v^{\si}(k)}{\Om_{p_{\ell'},\si}(k)}\, e^{-ik\cdot \nabla E_{p_{\ell'},\si}t}  a^*(k)  } \bigg)  h_t(\unp)\eta^*(\unp)^N\Ome,\quad\quad \label{LSZ-electrons}
\eeqa
where $h_t(\unp):=\prod_{\ell=1}^N \bigg(e^{-iE_{p_{\ell},\si}t  }h_{\ell}(p_{\ell})   \bigg)$ is the (renormalized) free evolution of  $h$. 

In the bracket in (\ref{LSZ-cloud}) we recognize the LSZ approximants of the clouds of real photons. For future reference we set
\beqa
\W_{p,\si}(t):=e^{\int d^3k\, \fr{v^{\si}(k)}{\Om_{p,\si}(k)}\, ( e^{-i|k|t}a^*(k)- e^{i|k| t}a(k))}. \label{cloud-of-photons}
\eeqa
It is more difficult to recast the expression in (\ref{LSZ-electrons}) as LSZ approximants pertaining to the electrons. For this purpose 
\emph{we reverse the Dollard prescription} in the expression  $e^{-ik\cdot \nabla E_{p,\si}t}$ in (\ref{LSZ-electrons}) that is we make a
substitution $e^{-ik\cdot \nabla E_{p_{\ell'},\si}t}\to e^{-ik\cdot x_{\ell}}$. 
This leads us to  the new family of approximating vectors
\beqa
& &\ti{\Psi}^{\si}_{h,t}=e^{iHt}\int d^{3N}\unp\,  e^{i\ga_{\unp,\si}(t)}e^{-\theta_{\unp,\si}(t) } \bigg(\prod_{\ell=1}^N  \W_{p_{\ell},\si}(t) \bigg)\times\non\\
& &\ph{44444444444444444444444}\times\bigg( \prod_{\ell'=1}^N e^{-D_{p_{\ell'},\si }} e^{-\int d^3k\, \fr{v^{\si}(k)}{\Om_{p_{\ell'},\si}(k)}\, e^{-ik\cdot x_{\ell'} t}  a^*(k)  } \bigg)  h_t(\unp)\eta^*(\unp)^N\Ome.\label{mass-renormalisation}\quad
\eeqa
Although we do not have a rigorous proof that  $\lim_{t\to\infty}\|\Psi^{\si}_{h,t}- \ti{\Psi}^{\si}_{h,t}\|=0$, it is intuitively clear, that the position $x$ of the freely evolving  electron   behaves asymptotically
as $\nabla E_{p,\si}t$. To simplify (\ref{mass-renormalisation}), we define the following (tentative) renormalized creation operator of the electron
\beqa
& &\ti \eta_{\si}^*(p):= \sum_{m=0}^{\infty}\fr{1}{\sqrt{m!}}\int d^{3m}k\,  \ti f^{m}_{p,\si}(k_1, \ldots, k_m) \bb^*(k_1)\ldots \bb^*(k_m)
\nn^*(p- \unk^{(m)}),\label{tentative-rco-zero}\\
& &\ti f^{m}_{p,\si}(k_1, \ldots, k_m):=(-1)^m e^{-D_{p,\si}} \fr{v^{\si}(k_1)}{\Om_{p,\si}(k_1)}\ldots \fr{v^{\si}(k_m)}{\Om_{p,\si}(k_m)},  
\label{actual-rco-zero}
\eeqa 
where $\unk^{(m)}=k_1+\cdots+k_m$.   Using $e^{-ik\cdot x}\eta^*(p)\Ome=\eta^*(p-k)\Ome$, it is then easy to show that
 \beqa
e^{-D_{p,\si}}\bigg(e^{-\int d^3k\, \fr{v^{\si}(k)}{\Om_{p,\si}(k)}\, e^{-ik\cdot x} a^*(k) } \bigg)\eta^*(p)\Ome=\ti \eta_{\si}^*(p)\Ome.
\eeqa
Thus, intuitively,  $\ti \eta_{\si}^*(p)$ creates from the vacuum the electron with its cloud of virtual photons.

Consequently, we can rewrite (\ref{mass-renormalisation}) in the LSZ form:
\beqa
\ti{\Psi}^{\si}_{h,t}=e^{iHt}\int d^{3N}\unp\,  e^{i\ga_{\unp,\si}(t)}e^{-\theta_{\unp,\si}(t) }\bigg( \prod_{\ell=1}^N  \W_{p_{\ell},\si}(t)\bigg)  \bigg(\prod_{\ell'=1}^N
e^{-iE_{p_{\ell'},\si}t}h_{\ell'}(p_{\ell'})  \ti\eta_{\si}^*(p_{\ell'})  \bigg)\Ome.  \label{final-formula-FK}
  \eeqa
The real-valued functions $\ga_{\unp,\si}$ and $\theta_{\unp,\si}$, appearing above, have the following explicit form  
\beqa
\ga_{\unp,\si}(t)\x&:=&\x {\ga}_{1;\unp,\si}(t)+{\ga}_{2;\unp,\si}(t),  \non\\
{\ga}_{1;\unp,\si}(t)\x &:= &\x -2  \sum_{\ell=1}^N\int d^3k\, v^{\si}(k)^2 \fr{ \sin\, \Om_{p_{\ell},\si}(k)t }{\Om_{p_{\ell},\si}^2(k)},   \label{phase-starts}\\
{\ga}_{2;\unp,\si}(t)\x&:=&\x-2\sum_{\ell<\ell'}     \int d^3k\, v^{\si}(k)^2   
\fr{(\sin  \Om_{p_{\ell'},\si}(k) t +\sin  \Om_{p_{\ell},\si}(k) t )}{ \Om_{p_{\ell},\si}(k) \Om_{p_{\ell'},\si}(k)} \label{mixed-phase} \\
& &+\sum_{\ell<\ell'}     \int d^3k\,   v^{\si}(k)^2   \bigg(\fr{1}{\Om_{p_{\ell},\si}(k)} +\fr{1}{ \Om_{p_{\ell'},\si}(k)} \bigg) \fr{\sin\,  (\Om_{p_{\ell},\si}(k) - \Om_{p_{\ell'},\si}(k))   t   }{  (\Om_{p_{\ell},\si}(k) - \Om_{p_{\ell'},\si}(k)) },  \label{Coulomb-phase}\\
\theta_{\unp,\si}(t)\x&:=&\x  \sum_{\ell<\ell'}\int d^3k \,v^{\si}(k)^2   \fr{ \cos(\Om_{p_{\ell'},\si}(k)-\Om_{p_{\ell},\si}(k) ) t}{ \Om_{p_{\ell},\si}(k)\Om_{p_{\ell'},\si}(k)}. \label{phase-ends} 
\eeqa
Recalling that $\Om_{p,\si}(k)=|k|-\nabla E_{p,\si}\cdot k$ and therefore $\Om_{p_{\ell},\si}(k)-\Om_{p_{\ell'},\si}(k)= (\nabla E_{p_{\ell'},\si}-\nabla E_{p_\ell,\si})\cdot k $
we expect that the above contributions facilitate the asymptotic decoupling between the following particles:
\begin{itemize}
\item (\ref{phase-starts}):  the $\ell$-th electron and a photon from the $\ell$-th cloud. 
\item (\ref{mixed-phase}): the $\ell$-th electron and  a photon from the $\ell'$-th cloud (and vice versa).
\item (\ref{Coulomb-phase}), (\ref{phase-ends}): the $\ell$-th electron and the $\ell'$-th electron.
\end{itemize}
Expression~(\ref{Coulomb-phase}) corresponds  to the Coulomb phase and it is easy to show that it behaves as $\log\, t$ for large $t$ and $\si=0$. The
remaining terms do not have counterparts in many-body quantum mechanical scattering.

\section{Comparison with a rigorous LSZ approach}\label{rigorous}   \setcounter{equation}{0}

For $N=1$ formula~(\ref{final-formula-FK}) is very similar to the single-electron state approximants obtained by Pizzo in \cite{Pi05}.
To obtain these latter states from (\ref{final-formula-FK}) one has to make the following modifications:

\begin{enumerate}

\vspace{0.2cm}

\item \nin\bf Cell partition: \rm The region of $p$-integration in (\ref{final-formula-FK}) has to be divided into time-dependent cubes.
Suppose, for convenience, that this region is a cube of volume equal to one, centered at zero. 
At time $1 \leq |t|$ the linear dimension of each cell is $1/2^{\np}$, 
where $\np\in\nat$ is s.t.
\beq
(2^{\np})^{1/\peps}\leq |t|< (2^{\np+1})^{1/\peps} \label{cell-partition-def}
\eeq 
for a small exponent $\peps>0$. Thus there are $2^{3\np}\leq |t|^{3\peps}$ cells.
Each such cell is denoted $\Ga^{(t)}_j$ and the collection of all cells $\Ga^{(t)}$. 

\item \nin\bf Photon clouds: \rm  The photon cloud $\W_{p,\si}(t)$ from  (\ref{final-formula-FK}) should be replaced with the cloud  
$\W_{\si}(\mrm{v}_j,t)$, defined in (\ref{cloud-W}) below, associated with the cube $\Ga_j^{(t)}$ containing $p$ and depending on the velocity  $\mrm{v}_j:=\nabla E_{p_j,\si}$  in the center of the cube $\Ga_j^{(t)}$. Thus one makes the following substitution
\beqa
& &\W_{p,\si}(t):=\exp\bigg\{-\int d^3k \, \vv^{\si}(k) \fr{a(k)e^{i|k|t}-a^*(k)e^{-i|k|t}}{|k|(1-\hatk\cdot {\bc \nabla E_{p,\si}} ) }   \bigg\} \label{cloud-zero}\\
& &\ph{444}\downarrow\non\\
& &\W_{\si}(v_j,t):=\exp\bigg\{-\int d^3k \, \vv^{\si}(k) \fr{a(k)e^{i|k|t}-a^*(k)e^{-i|k|t}}{|k|(1- \hatk\cdot {\bc \mrm{v}_j} ) }   \bigg\}, \label{cloud-W}
\eeqa
where $\mrm{v}_j:=\nabla E_{p_j,\si}$ is the velocity in the center of the cube $\Ga_j^{(t)}$ and $\hatk:=k/|k|$. Clearly, the difference  
$|\nabla E_{p,\si}-\mrm{v}_j|$ tends to
zero as $t\to \infty$ and the size of each cube $\Ga^{(t)}$ shrinks to zero, so it should not be difficult to justify this substitution.
\item \nin\bf Phases: \rm The phase $\ga_{p,\si}(t)$ from (\ref{final-formula-FK}) should be replaced with the phase defined in (\ref{slow-cutoff}) below.
Thus in view of (\ref{phase-starts}) and the definition above $\Om_{p, \si}(k):=|k|-k\cdot \nabla E_{p,\si}$, we make the substitution
\beqa
& &\ga_{p,\si}(t)=  -\int_{{\bc 0}}^t d\tau \bigg\{{\bc \int_{0\leq |k| }  d|k| } d\Omm(\hk)\, \vv^{\si}(k)^2(2|k|)
\bigg(\fr{\cos(k\cdot\nabla E_{p,\si}\tau-|k|\tau)}{1-\hatk\cdot {\bc \nabla  E_{p,\si} } }\bigg)  \bigg\} \label{tentative-slow-cut-off}  \\
& &\ph{444}\downarrow\non\\
& &\ga_{\si}(\mrm{v}_j,t)(p)=  -\int^t _{\bc 1} d\tau  \bigg\{ {\bc \int_{0\leq |k|\leq \si_{\tau}^\S} d|k| } d\Omm(\hk)\, \vv^{\si}(k)^2(2|k|)
\bigg(\fr{\cos(k\cdot\nabla E_{p,\si}\tau-|k|\tau)}{1-\hatk\cdot {\bc \mrm{v}_j } }\bigg)  \bigg\}, \label{slow-cutoff}
\eeqa
where  $d\Omm(\hat k):=\sin\,\theta_{\hat k} d\theta_{\hat k} d\phi_{\hat k}$ is the measure on the unit sphere,  and $\tau\mapsto \si^{\S}_{\tau}=\ka\tau^{-\al}$, $1/2<\al<1$, is the slow infrared cut-off.  (As stated in 5. below, the cut-off $\si$ will tend to zero with $t$ much faster).   Since the region of momenta 
$|k|\geq \si_{\tau}^{\S}$ affected by the above change  is well separated from the infrared singularity, it is easy to justify the above step using stationary phase arguments.

\item \nin\bf Renormalized creation operators: \rm The tentative renormalized creation operator of the electron (\ref{tentative-rco-zero})-(\ref{actual-rco-zero})
should be replaced with the actual  renormalized creation operator, given by (\ref{actual-rco}) below. That is, we make the following replacement:
\beqa
& &\ti \eta_{\si}^*(p):= \sum_{m=0}^{\infty}\fr{1}{\sqrt{m!}}\int d^{3m}k\,  {\bc \ti f^{m}_{p,\si}}(k_1, \ldots, k_m) \bb^*(k_1)\ldots \bb^*(k_m)
\nn^*(p- \unk^{(m)}), \label{tentative-rco}\\
& &\ph{444}\downarrow\non\\
& &\hat \eta_{\si}^*(p):= \sum_{m=0}^{\infty}\fr{1}{\sqrt{m!}}\int d^{3m}k\,  {\bc  f^{m}_{p,\si}}(k_1, \ldots, k_m) \bb^*(k_1)\ldots \bb^*(k_m)
\nn^*(p- \unk^{(m)}), \label{actual-rco}
\eeqa  
where the functions $\ti f^m_{p,\si}$ are given by (\ref{actual-rco-zero}) and  $f^{m}_{p,\si}$ are wave-functions of the normalized ground states $\psi_{p,\si}$
of the fiber Hamiltonians $H_{p,\si}$. These latter Hamiltonians are defined via the direct integral decomposition
\beqa
& &H^{(1)}_{\si}=\Pi^*\bigg(\int^{\oplus} d^3p\, H_{p,\si} \bigg)\Pi, \label{H-p}  \label{fiber-decomposition}
\eeqa
where $\Pi$ is a certain unitary identification of Hilbert spaces and $H_{p,\si}$ is a concrete operator on an auxiliary fiber Fock space $\mcF_{\fiber}$. 
The key property of the operator (\ref{actual-rco}) is that it creates a freely-evolving physical electron from the vacuum 
(at fixed $\si>0$), i.e.
\beqa
e^{iH_{\si}t}\int d^3p\, h(p)\hat\eta^*(p)\Ome=\Pi^*\int^{\oplus}d^3p\,  e^{-it E_{p,\si}t }h(p) \psi_{p,\si}. \label{free-evolution}
\eeqa
Starting from  \cite[formula~(4.43)]{DP17}, \cite[formula~(5.2)]{DP12} and using methods from these references one can show that
\beqa
f^{m}_{p,\si}(k_1,\ldots,k_m)=\ti{f}^{m}_{p,\si}(k_1, \ldots, k_m)+\cdots,
\eeqa
where the omitted terms are either  of order $\la$ or more regular near zero  than $\ti{f}^{m}_{p,\si}$, at least in some variables $k_i$. Thus in the weak coupling regime $\ti{f}^{m}_{p,\si}$ captures the leading part of the infrared singularity of   $f^{m}_{p,\si}$. Further analysis in this direction is needed to
justify the substitution (\ref{tentative-rco}) $\to$ (\ref{actual-rco}), which takes correlations between the virtual photons dressing the electron into account.

\item\textbf{Fast infrared cut-off:} The infrared cut-off $\si$ appearing in (\ref{final-formula-FK}) should be removed in the limit $t\to\infty$. More precisely, one sets 
\beqa
\si \rightarrow \si_t:= 1/t^{\be} ,
\eeqa
for $\be\geq 1$ sufficiently large.

\end{enumerate}

After the above changes, we obtain from (\ref{final-formula-FK}) the following approximating sequence
\beqa
\hat{\Psi}_{h,t}:=e^{iHt}\sum_{j\in \Ga^{(t)}  }   \W_{\si_t}(v_j,t)  \int_{\Ga^{(t)}_j} d^3p \,e^{-iE_{p,\si_t}t } e^{i   \ga_{\si_t}(v_j,t)(p) } h(p) \hat\eta_{\si_t}^*(p)\Ome.  \label{Pizzo-infraparticle}
\eeqa
It was rigorously proven by Pizzo in \cite{Pi05} that the outgoing and incoming single-electron states  $\hat\Psi_{h}^{\mathrm{in}/\mathrm{out} }:=\lim_{t\to -/+\infty} \hat{\Psi}_{h,t}$ exist  and are non-zero. 

Given the above considerations, there is hope for proving convergence of the Faddeev-Kulish type approximating sequence (\ref{scattering-states-approximants-x}) in the single-electron case by estimating the norm distance to the Pizzo state~(\ref{Pizzo-infraparticle}).   The most difficult parts will be 
the partial reversal of the Dollard prescription (\ref{LSZ-electrons}) $\to$ (\ref{mass-renormalisation}) and the step from the tentative to the actual renormalized creation operator of the electron (\ref{tentative-rco}) $\to$ (\ref{actual-rco}). A more ambitious  strategy consists in proving the existence of the limit of (\ref{dollard})
directly, e.g. via an application of the Cook's method. Also here it seems necessary to make contact with the renormalized creation operator 
$\hat\eta^*(p)$, in order to exploit the key property~(\ref{free-evolution}). We hope to come back to these problems in  future publications. 

So far there is no counterpart of the result of Pizzo  for two or more electrons. Actually, it is not even clear how the approximating sequence
(\ref{Pizzo-infraparticle}) should look like in this case. 
As  scattering of two electrons  in the Nelson model is currently under investigation \cite{DP12.0,DP12, DP17},  it is  worth pointing out that the Faddeev-Kulish type analysis from previous sections gives a  reasonable candidate. In fact, let us simply apply the modifications 1.--5. listed above
to the approximating vector (\ref{final-formula-FK}) in the case $N=2$. We obtain
\beqa
\hat{\Psi}^{(2)}_{h,t}\!\!&:=&\!\!e^{iHt}\sum_{j_1,j_2\in \Ga^{(t)}  }   \W_{\si_t}(v_{j_1},t) \W_{\si_t}(v_{j_2},t) 
\int_{\Ga^{(t)}_{j_1}\times \Ga^{(t)}_{j_2} } d^3p_1 d^3p_2\, e^{i\ga_{2;\un p,\si_t}(t) } e^{-\theta_{\unp,\si_t}(t) }\times \\
& &\times\bigg(e^{-iE_{p_1,\si_t}t } e^{i  \ga_{\si_t}(v_{j_1},t)(p_1)}   h_1(p_1) \hat\eta_{\si_t}^*(p_1) \bigg) \bigg( e^{-iE_{p_2,\si_t} t }e^{i\ga_{\si_t}(v_{j_2},t)(p_2) }   h_2(p_2) \hat\eta_{\si_t}^*(p_2)\bigg) \Ome,
\eeqa
where $\ga_{2;\un p,\si}$, $\theta_{\unp,\si}$ are given by (\ref{mixed-phase})-(\ref{phase-ends}) and may require some small modifications, akin to 
(\ref{tentative-slow-cut-off})$\to $(\ref{slow-cutoff}).  We are confident that the above observations will facilitate mathematically rigorous research on scattering of two electrons in the Nelson model.

\section{Conclusion}\label{conclusions}  \setcounter{equation}{0}

In this paper we revisited the Faddeev-Kulish approach to electron scattering  in the context of the massless Nelson model. 
In contrast to the original  paper of Faddeev and Kulish,  we applied the Dollard formalism according to the rules of the art, without dropping the lower boundary 
of integration. This led us to a  scattering matrix which is meaningful on the usual Fock space of free electrons, but does not commute with the total electron 
momentum. This latter point was clarified in the later part of our analysis, where we reformulated this scattering matrix in LSZ terms: 
The lower boundary of integration gives rise to  clouds of real photons which always carry some momentum.  
Furthermore, we checked that the resulting LSZ formula at the one-electron level reproduces  single-electron states constructed rigorously by Pizzo, up to minor
technical differences. Our observations provide clear-cut mathematical conjectures, which will facilitate rigorous research 
of  $N$-electron scattering in the massless Nelson model.  Our findings may also provide a more solid  basis for heuristic discussions of
scattering theory in QED, which is a popular topic in  current physics literature.

\vspace{0.5cm}

\nin\textbf{Acknowledgements:} I would like to thank A. Pizzo for numerous discussions on infrared problems and for explaining
his work \cite{Pi05}. Thanks are also due to B. Gabai for pointing out relevant references. Furthermore, I would like to 
thank the organizers of the `Wolfhart Zimmermann Memorial Symposium' at the MPI Munich and the organizers of 
the workshop `Infrared problems in QED 
and Quantum Gravity' at the Perimeter Institute, as the two events were essential for completing this paper. 
%
This work was  supported by the DFG within the Emmy Noether grant DY107/2-1.

\end{document}